\documentclass[aps,superscriptaddress,twocolumn,amsmath,amssymb,nofootinbib]{revtex4}

\usepackage{color}
\usepackage{graphicx}
\usepackage{bm}
\definecolor{Blue}{rgb}{0.00, 0.00, 1.00}
\definecolor{Red}{rgb}{1.00, 0.00, 0.00}

\begin{document}

\title{Semi-analytical model of the contact resistance in two-dimensional semiconductors}

\author{Roberto Grassi}
\email{rgrassi@umn.edu}
\affiliation{Department of Electrical and Computer Engineering, University of Minnesota, 200 Union St. SE, Minneapolis, MN 55455, USA}
\author{Yanqing Wu}
\affiliation{Wuhan National High Magnetic Field Center and School of Optical and Electronic Information, Huazhong University of Science and Technology, Wuhan 430074, China}
\author{Steven J. Koester}
\author{Tony Low}
\affiliation{Department of Electrical and Computer Engineering, University of Minnesota, 200 Union St. SE, Minneapolis, MN 55455, USA}

\date{\today}
\begin{abstract}
Contact resistance is a severe performance bottleneck for electronic devices based on two-dimensional layered (2D) semiconductors, whose contacts are Schottky rather than Ohmic. Although there is general consensus that the injection mechanism changes from thermionic to tunneling with gate biasing, existing models tend to oversimplify the transport problem, by neglecting the 2D transport nature and the modulation of the Schottky barrier height, the latter being of particular importance in back-gated devices. In this work, we develop a semi-analytical model based on Bardeen's transfer Hamiltonian approach to describe both effects. Remarkably, our model is able to reproduce several experimental observations of a metallic behavior in the contact resistance, i.e., a decreasing resistance with decreasing temperature, occurring at high gate voltage.
\end{abstract}
\maketitle

\emph{Introduction---} 2D layered semiconducting materials, such as transition metal dichalcogenides (TMDs) and black phosphorus (BP), have many interesting electrical and optical properties \cite{novoselov2005two,wang2012electronics,xu2014spin,fiori2014electronics,low2016polaritons,koppens2014photodetectors,sun2016optical}, but tend to form Schottky barriers (SB) at the interfaces with metal contacts, resulting in a large contact resistance that severely degrades the device performance \cite{das2013does,du2014device,haratipour2015black}.

Thermionic emission \cite{sze2006physics} is commonly assumed when extracting the SB height from temperature-dependent current measurements of field-effect transistors (FETs). In \cite{allain2015electrical}, it was pointed out that this procedure is correct only at the gate voltage corresponding to the flat-band condition. Considering an n-type device, for example, above the flat-band voltage, the conduction band edge in the channel is higher than at the interface with the contact, hence electrons traversing the channel see a larger barrier than the SB height. Below the flat-band voltage, tunneling starts to contribute and the thermionic emission theory loses validity. As a result, this can lead to unphysical negative SB heights \cite{yu2014graphene,avsar2015air}. Furthermore, experiments show that, as opposed to the insulating behavior of a SB contact, the two-terminal resistance \cite{liu2015toward} as well as contact resistance \cite{cui2015multi} can decrease with decreasing temperature at high gate voltage. The origin of this metallic behavior is debated and not yet clarified \cite{radisavljevic2013mobility,allain2015electrical}.

\begin{figure}[t]
\centering
\scalebox{0.55}[0.55]{\includegraphics*[viewport=40 360 470 940]{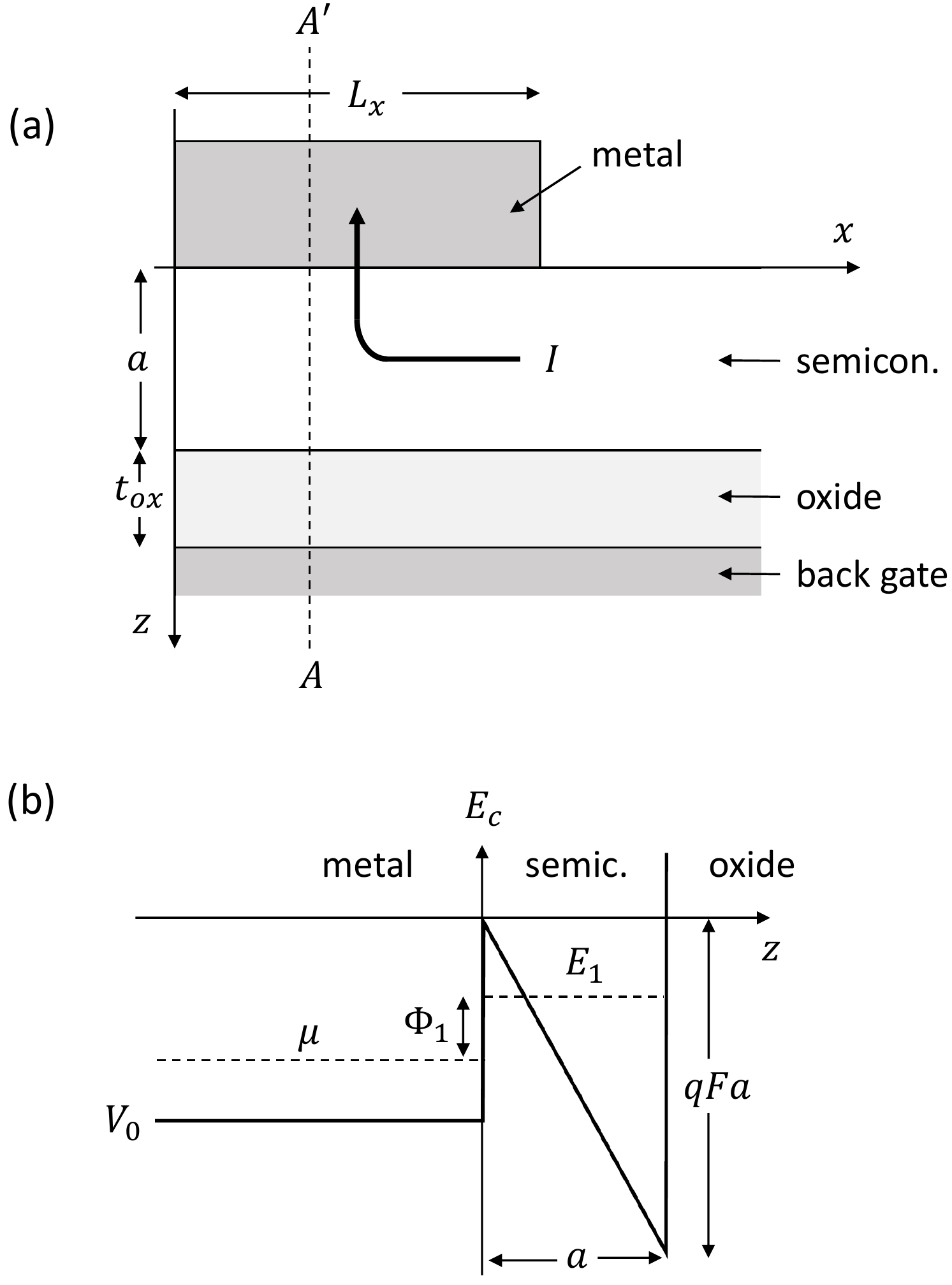}}
\caption{(a) Cross-section of the device structure under consideration. The path of the current flow is also indicated schematically. (b) Triangular barrier model of the conduction band edge profile along the cut line $A$--$A'$ in (a). The top of the barrier is taken as the energy reference. $F$ is the vertical electric field in the semiconductor. The SB height $\Phi_1$ is the energy difference between the first subband edge $E_1$ in the semiconductor and the Fermi level of the metal $\mu$. 
}
\label{fig_device}
\end{figure}

Recently, a model for SB FETs has been proposed in \cite{penumatcha2015analysing} and applied to extract the SB height and bandgap of BP devices. This model assumes one-dimensional transport and a bias-independent SB height. However, in a typical geometry with a top contact to a multilayer 2D semiconductor as in the sketch of Fig.~\ref{fig_device}a, transport is inherently 2D. To be precise, due to quantization, the SB height $\Phi_1$ to a 2D semiconductor should be defined as the difference between the edge $E_1$ of the first energy subband in the semiconductor and the Fermi level of the metal $\mu$ (see the schematic band profile in Fig.~\ref{fig_device}b for an n-type device). Transport occurring at energies above (below) $E_1$ is generally referred to as ``thermionic'' (``tunneling''). However, in the presence of a back gate, the subband edge and thus the SB height are expected to be modulated by the vertical electric field. When $E_1$ is lower than the bulk band edge at the interface with the metal, even the electrons traversing the junction at energies above $E_1$ see a tunneling barrier in the vertical (i.e., $z$) direction and a new transport regime arises.

In this paper, we present a semi-analytical model of the contact resistance to multilayer 2D semiconductors in this ``vertical tunneling'' regime. The model is based on a triangular barrier approximation of the vertical potential profile in the semiconductor underneath the contact. 2D transport is separated into a sequence of two 1D mechanisms: (\textit{i}) quantum tunneling through the SB at the metal-to-semiconductor interface, followed by (\textit{ii}) semiclassical ``diffusive'' transport across the semiconductor (the source of scattering being the in- and out-tunneling across the SB). The model is benchmarked against numerical solutions of the 2D quantum transport problem and employed to study the dependence of the contact resistance on vertical electric field and temperature. We show that, when the SB height is sufficiently lowered by the vertical electric field, contact resistance shows a metallic behavior with temperature, as observed in experiments. The model predicts a smooth transition from a thermionic-like regime at low electric field, where the tunneling barrier is almost transparent, to a true vertical tunneling regime at high electric field. In the former case, the extraction method of the SB height based on thermionic emission theory can still be applied.

\emph{Model---} We consider a single planar junction of length $L_x$ between a metal and a multi-layer 2D semiconductor with thickness $a$ (Fig.~\ref{fig_device}a). A vertical electric field is created inside the semiconductor by the presence of a back gate. Let $x$ and $z$ be the longitudinal and vertical directions, respectively. The device is uniform in the $y$ direction. We focus on the portion of the semiconductor covered by the metal (contact region) assuming that the uncovered part (channel) simply acts as a ``reflectionless'' contact or semi-infinite lead \cite{datta1997electronic}. A current can flow as indicated in Fig.~\ref{fig_device}a.

We neglect hole transport and discuss only injection of electrons from the metal to the conduction band of the semiconductor. A simple single-valley effective mass Hamiltonian, with the same values of the effective masses $m_{x,y,z}$ in the metal and in the semiconductor, is adopted. Such effective mass model has been shown to provide an accurate description of the out-of-plane quantization in multilayer BP \cite{zhang2017infrared}. For the case of multilayer TMDs, one should employ a more fundamental tight-binding model \cite{kang2016unified}. Whereas the form of the equations will be different, the general trends are not expected to change significantly. The bulk band edge profile is assumed to be uniform along the $x$ direction and is approximated with a triangular barrier along the $z$ direction, as shown in Fig.~\ref{fig_device}b, where the value of the band edge in the metal $V_0$ is chosen low enough so that it provides significant density of states at the Fermi level. Within this non-self-consistent approximation, which is valid at low carrier concentration, the magnitude of the vertical electric field $F$ in the semiconductor is simply proportional to the voltage $V_G$ applied between the back gate and the top metal:
\begin{equation}
F = \frac{V_G}{a + (\epsilon_s/\epsilon_\mathrm{ox})t_\mathrm{ox}} \, ,
\end{equation}
where the workfunctions of the two metals are taken to be equal, $t_\mathrm{ox}$ is the back oxide thickness, and $\epsilon_s$ and $\epsilon_\mathrm{ox}$ are the dielectric constants of the semiconductor and oxide, respectively. 

Due to vertical confinement, the energy spectrum in the semiconductor splits into a set of discrete 2D subbands. Within a triangular well approximation, the subband edges $E_i$ ($i$ positive integer) can be computed as \cite{miller2008quantum}
\begin{equation}
E_i = - qF \left( a - \frac{| \zeta_i |}{k_F} \right) , \label{eq_Ez}
\end{equation}
where the energy reference is taken at the top of the barrier in Fig.~\ref{fig_device}b, $q$ is the elementary electric charge, the wavevector $k_F$ is defined as
\begin{equation}
k_F = \left( \frac{2 m_z qF}{\hbar^2} \right)^{1/3} ,
\end{equation}
and $\zeta_i$ are the zeros of Airy's function, i.e., $\mathrm{Ai}(\zeta_i)=0$, which can be approximated as \cite{abramowitz1964handbook}
\begin{equation} \label{eq_zeros}
\zeta_i \approx - \left[ \frac{3 \pi}{8} (4 i-1) \right]^{2/3} .
\end{equation}
We limit the discussion to the case $E_i<0$. Indeed, the subband description looses validity above the barrier.

We assume that transport within the semiconductor can be described by a set of decoupled 1D Boltzmann's transport equations \cite{rudan2015physics}, one for each subband, where the tunneling from the metal to the semiconductor and vice versa is included as a scattering mechanism. The corresponding relaxation time $\tau_i$, or inverse of the probability rate that an electron originally in the $k$-space state $(k_x,k_y)$ of the $i$-th subband tunnels into the metal, is computed according to Bardeen's transfer Hamiltonian theory \cite{bardeen1961tunnelling,harrison1961tunneling,duke1969tunneling}, which has been recently applied to describe tunneling in vertical heterostructures of 2D materials \cite{feenstra2012single,britnell2013resonant} and electron-hole bilayer tunnel FETs \cite{alper2013quantum,agarwal2014engineering}. In the limit of large $L_x$, we get
\begin{equation} \label{eq_relaxation_time}
\frac{1}{\tau_i} = \left\{
\begin{array}{ll}
\frac{1}{h} \frac{\hbar^2}{2 m_z} \frac{k_F^2}{\mathrm{Ai}'^2(\zeta_i)} \frac{4 \sqrt{- E_i (E_i - V_0)}}{-V_0} e^{-2\gamma_0}, & \, V_0 < E_i < 0 \\
0, & \, \text{otherwise}
\end{array}
\right.
\end{equation}
where $h$ is Planck's constant, $\hbar=h/(2\pi)$, $\gamma_0$ is defined as
\begin{align}
\gamma_0 &= \frac{2}{3} \zeta_0^{3/2} \, , \\
\zeta_0 &= - k_F \frac{E_i}{qF} = k_F a - | \zeta_i | \, ,
\end{align}
and $\mathrm{Ai}'(\zeta_i)$ is the derivative of Airy's function evaluated at $\zeta_i$, which can be approximated as \cite{abramowitz1964handbook}
\begin{equation} \label{eq_derivative_at_zeros}
|\mathrm{Ai}'(\zeta_i)| \approx \frac{1}{\sqrt{\pi}} \left[ \frac{3 \pi}{8} (4 i-1) \right]^{1/6} .
\end{equation}
Note that $\tau_i$ is independent of both $k_x$ and $k_y$. The tunneling current is computed from the $x$-dependent distribution function of each subband, which is obtained by solving Boltzmann's transport equation with $\tau_i$ as the scattering relaxation time and with appropriate boundary conditions. In particular, we assume that the electrons are backscattered at the left end of the contact region. Differentiating the tunneling current with respect to the applied bias $V_D$ and taking the limit $V_D \ll k_B T/q$ ($k_B$ is Boltzmann's constant and $T$ the temperature) gives us the low bias conductance $G$ or inverse of contact resistance (per unit width). We obtain the semi-analytical expression
\begin{align} 
G & = \frac{2q^2}{h} \int_{-\infty}^{\infty} d \varepsilon \, \overline{T}(\varepsilon) \left(-\frac{\partial F_0}{\partial \varepsilon}\right) , \label{eq_cond} \\
F_0(\varepsilon) &= \sqrt{\frac{m_y k_B T}{2 \pi \hbar^2}} \mathcal{F}_{-1/2}\left(\frac{\mu - \varepsilon}{k_B T}\right) ,
\end{align}
where $\varepsilon$ is the total energy for electrons with $k_y=0$ and $\mathcal{F}_{-1/2}$ the Fermi-Dirac integral of order $-\tfrac{1}{2}$. The total transmission function $\overline{T}$ is defined as
\begin{equation} \label{eq_transmission_function}
\overline{T}(\varepsilon) = \sum_i T_i(\varepsilon) \, ,
\end{equation}
with the trasmission probability $T_i$ of each subband given by
\begin{equation} \label{eq_transmission}
T_i(\varepsilon) = \left\{
\begin{array}{ll}
0, & \quad \varepsilon < E_i \\
1 - e^{-\frac{2 L_x}{\lambda_i}}, & \quad \varepsilon > E_i 
\end{array}
\right. \, ,
\end{equation}
where $\lambda_i = |v_x| \tau_i$ is the mean free path related to tunneling and $|v_x| = \sqrt{ 2 (\varepsilon-E_i)/m_x }$ is the longitudinal carrier velocity. In the Supporting information, we provide a detailed derivation of the model.

\begin{figure}[t]
\centering
\scalebox{0.55}[0.55]{\includegraphics*[viewport=20 280 400 900]{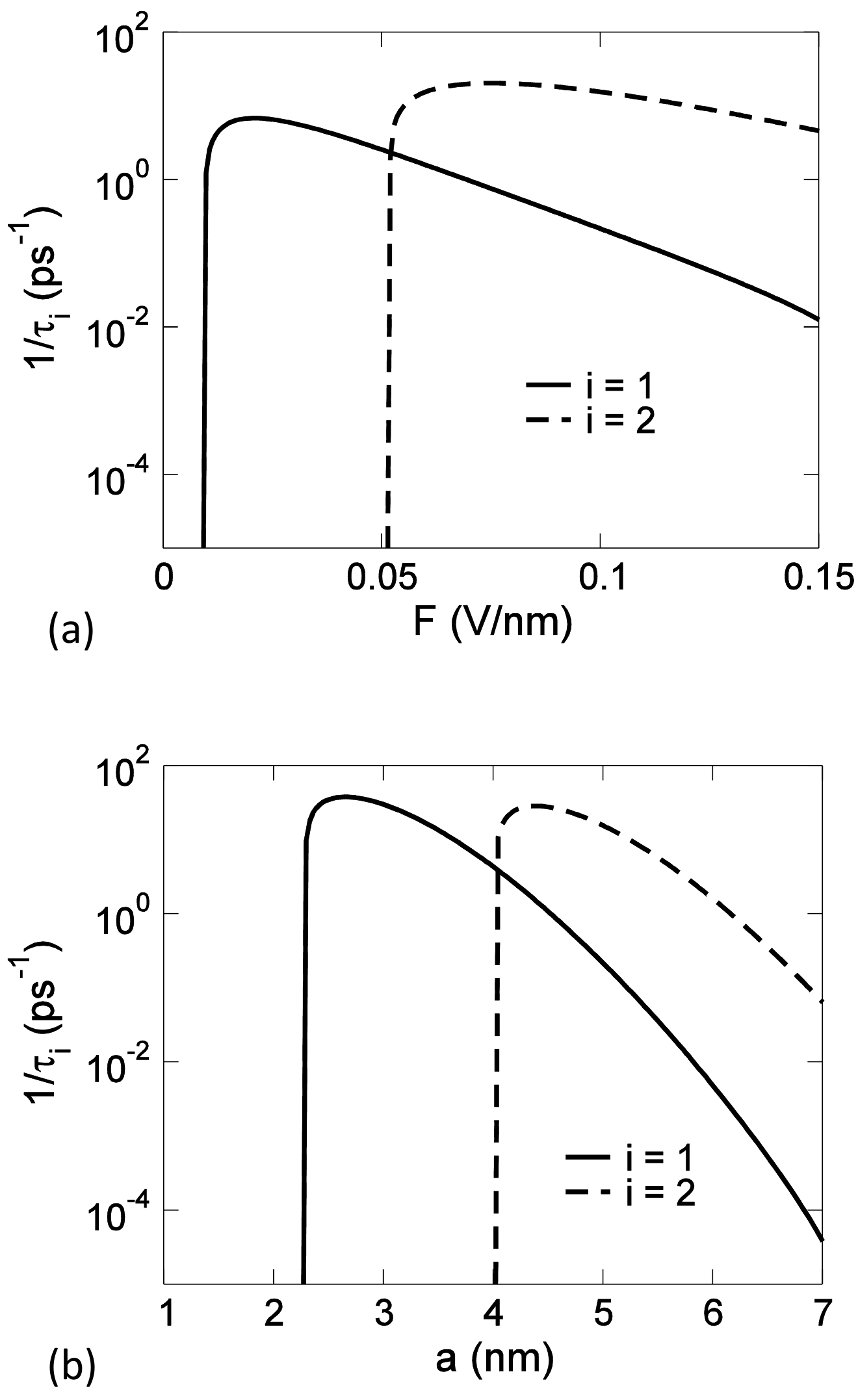}}
\caption{Tunneling rate of the first ($i=1$) and second ($i=2$) subband, computed from (\ref{eq_relaxation_time}) and plotted as a function of electric field $F$ at fixed semiconductor thickness $a=5$~nm in (a), and as a function of $a$ at fixed $F=0.1$~V/nm in (b). The other parameter values are: $m_z=0.4m_0$ ($m_0$ is the free electron rest mass) and $V_0=-0.5$~eV.
}
\label{fig_tauinv}
\end{figure}

\emph{Results---} In Fig.~\ref{fig_tauinv} we plot the tunneling rate $1/\tau_i$ of the first two subbands as a function of electric field and semiconductor thickness. The predicted tunneling rate goes to zero at small $F$ or $a$ because the Bardeen model does not account for the above-the-barrier regime at $E_i>0$. At high electric field or large semiconductor thickness, the exponential term in (\ref{eq_relaxation_time}) is dominating. In this regime, an increase of $F$ or $a$ results in a decrease of the scattering rate $1/\tau_i$. This can be understood by noting that, since both $k_x$ and $k_y$ are conserved in the tunneling process, an electron can tunnel from the metal to the semiconductor only if its vertical energy is equal to $E_i$. However, according to (\ref{eq_Ez}), $E_i$ shifts to lower energies with increasing $F$ or $a$. Because of that shift, the tunneling distance, which is equal to $|E_i|/(qF)$ at the vertical energy $E_i$, becomes longer as $F$ or $a$ increase.

\begin{figure*}[t]
\centering
\scalebox{0.55}[0.55]{\includegraphics*[viewport=20 190 950 440]{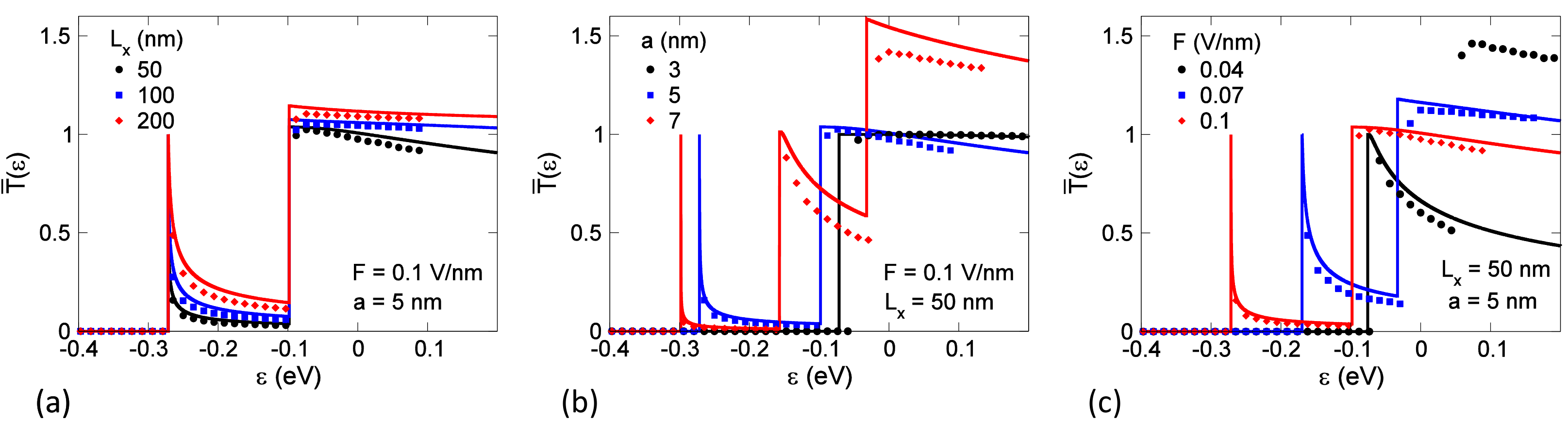}}
\caption{Comparison between the transmission function vs. energy from the model in (\ref{eq_transmission}) (lines) and from GF (symbols) for different values of (a) contact length $L_x$, (b) semiconductor thickness $a$, and (c) electric field $F$. Other parameter values are: $m_x=0.2m_0$, $m_z=0.4m_0$, $V_0=-0.5$~eV. 
}
\label{fig_tran}
\end{figure*}

In order to benchmark the proposed model, we solve numerically the 2D Schr\"odinger equation with open boundary conditions using the Green function (GF) method \cite{datta1997electronic} and assuming the same non-self-consistent triangular potential profile as in Fig.~\ref{fig_device}b. Details on the GF calculation can be found in the Supporting information. Fig.~\ref{fig_tran} shows the plot of the total transmission function $\overline{T}(\varepsilon)$ computed with the analytical expression in (\ref{eq_transmission_function})--(\ref{eq_transmission}) and with GF for different sets of parameter values. The two models are in good general agreement. The transmission function increases by one at each energy corresponding to a subband edge $E_i$, indicating a resonant tunneling regime, and shows a decaying behavior between two successive subband edges. Indeed, different energies correspond to different $k_x$ states. Since the length of the contact $L_x$ is finite and the transfer length or average distance traveled by an electron in the semiconductor before tunneling into the metal is equal to the mean free path $\lambda_i = |v_x| \tau_i$, the probability of escaping into the metal is larger for the states closer to the subband edge which have smaller velocity. As shown in Fig.~\ref{fig_tran}, increasing the contact length tends to raise the transmission probability of each single subband to unity because the ratio $L_x/\lambda_i$ between the contact length and the average distance before tunneling increases, which means that the electrons have more chances to enter the contact. The shift of $E_i$ to lower energy as $a$ or $F$ increase is clearly seen from the shift of the transmission peaks in Fig.~\ref{fig_tran}b and c, respectively. This is accompanied by a narrowing of the peaks, which is related to the decrease of $1/\tau_i$ discussed above.

It should be noted that, with reference to the generic subband of index $i$, our model predicts no vertical tunneling contribution at energies below $E_i$ (see Eq.~\ref{eq_transmission}). This process could be possible in the case of a realistic band bending between the channel and the contact region. However, since tunneling is decreasing exponentially with the tunneling distance, the effect would be concentrated at the contact edge. Therefore, if $E_i$ is sufficiently close to the metal Fermi level $\mu$, the contribution from energies below $E_i$ (lateral tunneling) is negligible compared to energies above $E_i$ (vertical tunneling) because of the large surface-to-edge ratio of the contact.

\begin{figure*}[t]
\centering
\scalebox{0.55}[0.55]{\includegraphics*[viewport=20 30 710 520]{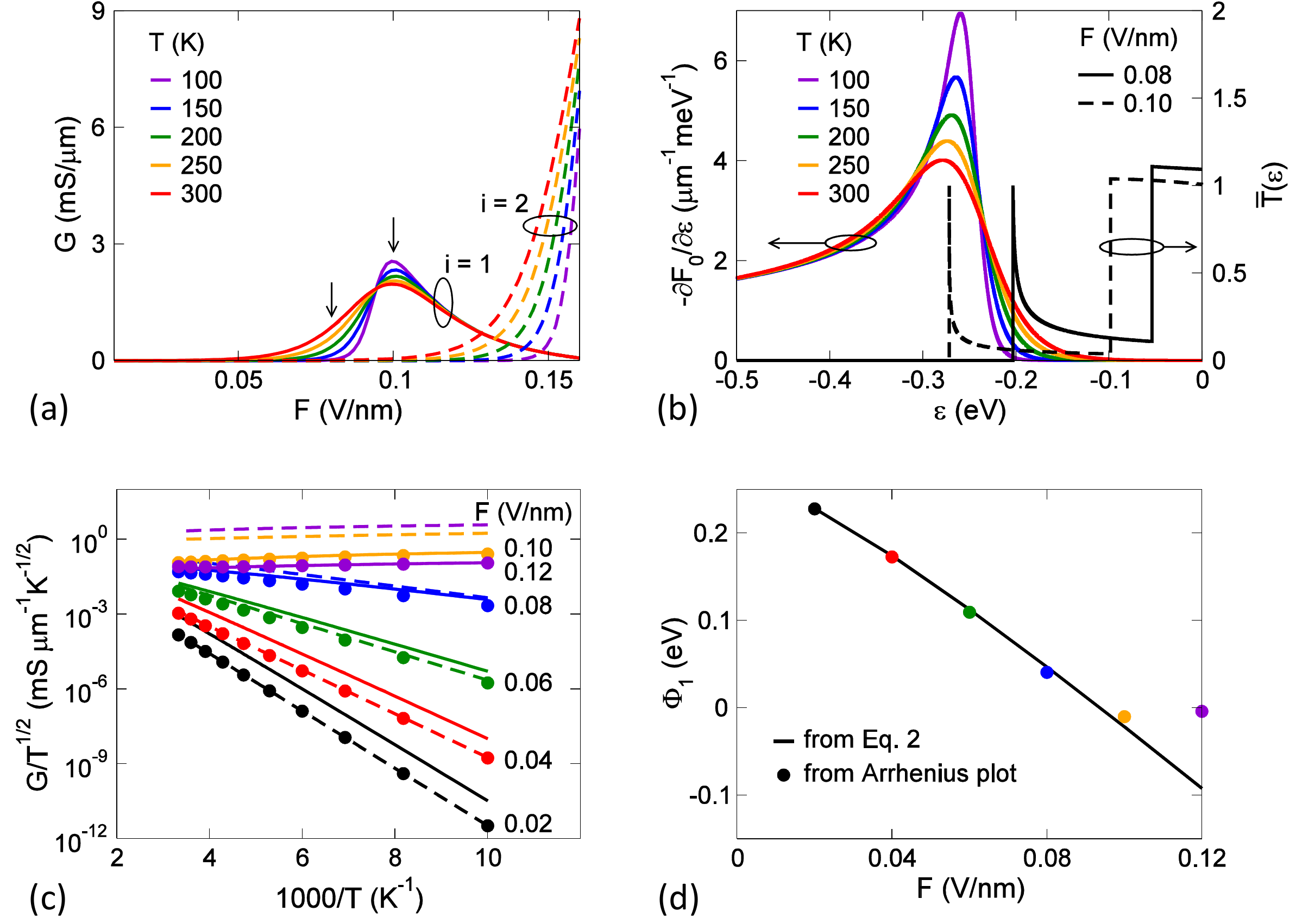}}
\caption{(a) Conductance $G$ vs. vertical electric field for different temperatures at $\mu=-0.25$~eV. The contributions of the first ($i=1$) and second ($i=2$) subband are separated. (c) Plot of $-\partial F_0/\partial \varepsilon$ vs. energy for different temperatures at $\mu=-0.25$~eV. Superimposed are the spectra of the transmission function at the two electric field values indicated by arrows in (a). (c) Arrhenius plot of $G/T^{1/2}$ at different electric field values, computed from the rigorous model in (\ref{eq_cond}) (symbols) and the approximated expressions in (\ref{eq_cond_approx1}) (dashed lines) and (\ref{eq_cond_approx2}) (solid lines). (d) SB height $\Phi_1$ vs. electric field, extracted from the average slope of the Arrhenius plot with symbols in (c) assuming thermionic emission (symbols), compared with the actual values from (\ref{eq_Ez}) (line). Other parameters values are: $m_x=0.2m_0$, $m_y=m_0$, $m_z=0.4m_0$, $V_0=-0.5$~eV, $a=5$~nm, $L_x=50$~nm.
}
\label{fig_cond}
\end{figure*}

Fig.~\ref{fig_cond}a plots $G$, obtained by numerically computing the energy integral in (\ref{eq_cond}) and resolved for the first two subbands, as a function of $F$ and temperature $T$. It can be seen that $G$ is a non-monotonic function of the electric field. This is consistent with our previous observations: the transmission probability reduces with increasing $F$ because the tunneling distance increases. Fig.~\ref{fig_cond}a shows that, at high electric field before the second subband starts to contribute significantly, the derivative $\partial G/\partial T$ is negative, i.e., contact resistance decreases with decreasing temperature. This has to do with the factor $-\partial F_0/\partial \varepsilon$, which is plotted in Fig.~\ref{fig_cond}b. It can be proved that its derivative with respect to temperature changes sign at the energy $\varepsilon_0 \approx \mu+0.857 k_B T$, which has only a weak temperature dependence ($\varepsilon_0=-0.243$ and $-0.228$~eV at $T=100$ and $300$~K, respectively) and appears as a crossover point in Fig.~\ref{fig_cond}b. As the transmission function shifts to lower energy with increasing $F$ (compare the plots at $F=0.08$ and $0.1$~V/nm in Fig.~\ref{fig_cond}b), more contribution to the integral in (\ref{eq_cond}) comes from the energy range where $\partial(-\partial F_0/\partial \varepsilon)/\partial T < 0$ and eventually leads to $\partial G/\partial T<0$.

The model in (\ref{eq_cond}) allows for an analytical solution in two limiting cases. In order to simplify the discussion, we assume that only the first subband contributes to transport. Note that the energy range relevant for transport goes from $E_1$ to few $k_B T$'s above $E_1$ or $\mu$, whichever is maximum. If $L_x \gg \lambda_1$ in this energy range, it follows that $T_1(\varepsilon)\approx 1$ (i.e., an almost transparent barrier) and (\ref{eq_cond}) simplifies to
\begin{equation} \label{eq_cond_approx1}
G \approx \frac{2q^2}{h} F_0(E_1) \, .
\end{equation}
If, in addition, $\Phi_1=E_1-\mu \gg k_B T$, then (\ref{eq_cond_approx1}) further reduces to
\begin{equation} \label{eq_cond_thermionic}
G \propto  T^{1/2} \exp\left( -\frac{\Phi_1}{k_B T} \right) ,
\end{equation}
which is the expression of the thermionic emission theory for a 2D system \cite{anwar1999effects}. The prefactor is $T^{1/2}$ instead of $T^{3/2}$ because we are considering the low bias limit $V_D \ll k_B T/q$. Expression (\ref{eq_cond_thermionic}) implies that the SB height $\Phi_1$ can be extracted from the slope of $\ln(G/T^{1/2})$ vs $1/T$. On the other hand, if $L_x \ll \lambda_1$ in most of the energy window for transport, one can derive (see Supporting information)
\begin{equation} \label{eq_cond_approx2}
G \approx \frac{2q^2}{h} \frac{\sqrt{m_x m_y}}{\hbar} \frac{L_x}{\tau_1} f_0(E_1) \, ,
\end{equation}
where $f_0(E) = \{\exp[(E-\mu)/(k_B T)]+1\}^{-1}$ is the Fermi-Dirac function. Fig.~\ref{fig_cond}c compares the Arrhenius plot of $G/T^{1/2}$ computed with the rigorous model in (\ref{eq_cond}) and the approximated expressions in (\ref{eq_cond_approx1}) and (\ref{eq_cond_approx2}). $E_1$ is calculated according to (\ref{eq_Ez}) in all three cases. Similar Arrhenius plots are commonly used to extract the SB height in experiments \cite{yu2014graphene,anugrah2015determination,avsar2015air,cui2015multi}. For the chosen set of parameter values, approximation (\ref{eq_cond_approx1}) is valid up to $F\approx 0.08$~V/nm. At higher electric field, the transmission function becomes increasingly peaked around the subband edge (see Fig.~\ref{fig_cond}b) and (\ref{eq_cond_approx2}) becomes a better approximation. A positive slope, or metallic behavior, is predicted at high electric field similar to what has been reported in experiments \cite{yu2014graphene,avsar2015air}. In Fig.~\ref{fig_cond}d, we plot the SB height obtained by fitting the data of the rigorous model in (\ref{eq_cond}) with the thermionic expression (\ref{eq_cond_thermionic}), compared with the actual value of $\Phi_1=E_1-\mu$ from (\ref{eq_Ez}). It is seen that the extraction method based on the thermionic emission theory can provide good results at low electric field values, where the tunneling barrier is almost transparent. In the high-field regime, a fitting based on (\ref{eq_cond_approx2}) would provide a more physical result.

We conclude by noting that the model presented in this work can be easily extended to account for a finite carrier mobility in the semiconductor by introducing an additional relaxation time $\tau_s$ (and a corresponding mean free path $\lambda_s = |v_x| \tau_s$) related to elastic scattering. The main effect of scattering would be, for each subband, a shorter transfer length and a transmission probability that does saturate to unity in the limit of a long contact length. In the regime when only one subband is populated, the model could also be extended to include self-consistent electrostatics using the variational approach in \cite{stern1972self}.

\emph{Conclusions---} In summary, we have demonstrated that the metallic behavior of the contact resistance observed in recent experiments can be explained by taking into account the modulation of the vertical tunneling due to the SB lowering with increasing electric field in back-gated devices. To the best of our knowledge, this transport regime has not been discussed before. The model also suggests a non-monotonic behavior of the contact resistance with respect to vertical electric field and semiconductor thickness. Our semi-analytical model provides a reasonable description of contact resistance in 2D semiconductors and could be useful for contact engineering in future 2D electronics.

\onecolumngrid
\clearpage

\begin{center}
\textbf{\large Supporting information: Semi-analytical model of the contact resistance in two-dimensional semiconductors}
\end{center}
\setcounter{equation}{0}
\setcounter{figure}{0}
\setcounter{table}{0}
\setcounter{page}{1}
\makeatletter
\renewcommand{\theequation}{S\arabic{equation}}
\renewcommand{\thefigure}{S\arabic{figure}}
\renewcommand{\bibnumfmt}[1]{[S#1]}
\renewcommand{\citenumfont}[1]{S#1}

\section{Model of vertical tunneling}

\begin{figure}[b]
\centering
\scalebox{0.55}[0.55]{\includegraphics*[viewport=30 190 940 400]{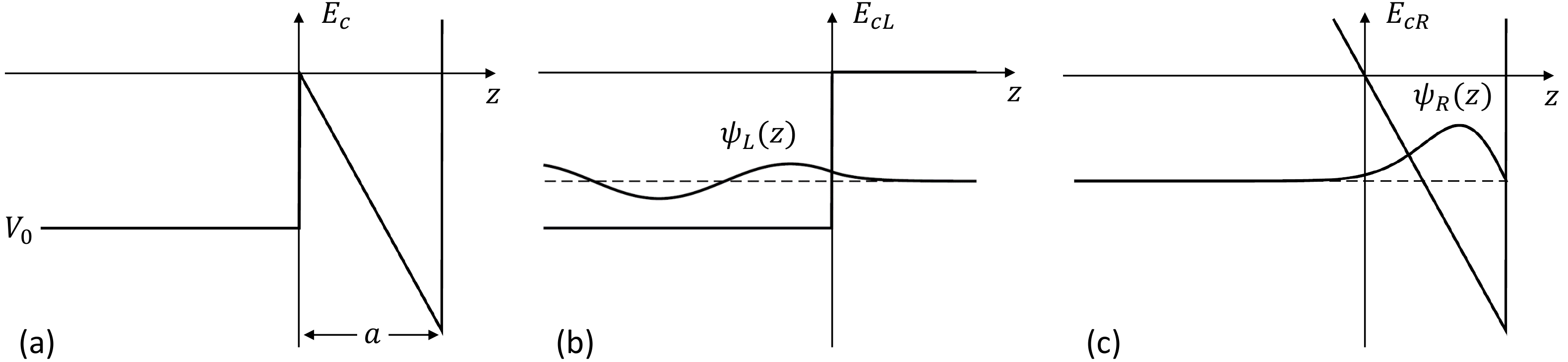}}
\caption{Energy band profiles of the different Hamiltonians: (a) $E_c(z)$, (b) $E_{cL}(z)$, (c) $E_{cR}(z)$. The wavefunctions corresponding to the cases (b) and (c) for motion normal to the junction are also represented schematically.
}
\label{fig_potential}
\end{figure}
We compute the tunneling rate in the limit of an infinite contact length and assume that the electric potential does not depend on the longitudinal position $x$, which implies translational invariance along $x$. Let $z=0$ be the vertical position of the metal-to-semiconductor interface. We consider a simple effective mass Hamiltonian
\begin{equation} \label{eq_hamiltonian}
\mathcal{H} = -\frac{\hbar^2}{2} \nabla \cdot \hat{m}^{-1} \nabla + E_c(z) = \mathcal{T} + E_c(z), \quad
\hat{m} = \left(
\begin{array}{ccc}
m_x & 0 & 0 \\
0 & m_y & 0 \\
0 & 0 & m_z
\end{array}
\right),
\end{equation}
where the values of the effective masses $m_{x,y,z}$ are taken to be the same in the metal and in the semiconductor and the conduction band edge profile $E_c(z)$ is modeled as a triangular barrier ($F > 0$ is the magnitude of the vertical electric field, see Fig.~\ref{fig_potential}a):
\begin{equation}
E_c(z) = \left\{ \label{eq_potential}
\begin{array}{ll}
V_0, & \quad z< 0 \\
- q F z, & \quad 0 <z< a \\
\infty, & \quad z > a
\end{array}
\right.
\end{equation}

Note that this is different from the Fowler-Nordheim field-emission problem \cite{fowler1928electron} because of the presence of the hard wall at $z=a$. Suppose that an electron is launched from $z<0$ towards the interface. The electron wavefunction will be totally reflected at $z=a$, resulting in a reflection coefficient, measured as the ratio between the probability currents of reflected and incident waves, identically equal to one at all energies. Does it mean that the tunneling probability is zero?  The Bardeen Transfer Hamiltonian method \cite{bardeen1961tunnelling_S,harrison1961tunneling_S,duke1969tunneling_S} provides a way to overcome this difficulty: the tunneling process across the barrier is thought of as a scattering event between states localized on different sides of the junction and the corresponding transition probability is computed through Oppenheimer's version of time-dependent perturbation theory \cite{oppenheimer1928three}. 

More precisely, $\mathcal{H}$ is taken as the perturbed Hamiltonian acting in the time interval $0<t<t_P$. For $t<0$, the unperturbed Hamiltonian must be identified with an Hamiltonian $\mathcal{H}_L$ that approximates well the true Hamiltonian $\mathcal{H}$ on the metal side of the junction but whose eigenfunctions decay in the semiconductor. We take $\mathcal{H}_L = \mathcal{T} + E_{cL}(z)$ with $E_{cL}(z)$ a potential step (Fig.~\ref{fig_potential}b):
\begin{equation} \label{eq_potential_left}
E_{cL}(z) = \left\{
\begin{array}{ll}
V_0, & \quad z< 0 \\
0, & \quad z > 0
\end{array}
\right.
\end{equation}
For $t>t_{P}$, one must choose a different unperturbed Hamiltonian $\mathcal{H}_R$ which, conversely, approximates well $\mathcal{H}$ on the semiconductor side of the junction but whose eigenfunctions decay in the metal. We take $\mathcal{H}_R = \mathcal{T} + E_{cR}(z)$ with $E_{cR}(z)$ a triangular well (Fig.~\ref{fig_potential}c):
\begin{equation}
E_{cR}(z) = \left\{ \label{eq_potential_right}
\begin{array}{ll}
- q F z, & \quad z < a \\
\infty, & \quad z > a
\end{array}
\right.
\end{equation}
It is assumed that prior to the perturbation the electron wavefunction coincides with an eigenfunction $\psi_{L,\alpha}$ of $\mathcal{H}_L$. Since this is not an eigenstate of $\mathcal{H}$, the electron wavefunction will evolve during the time interval $0<t<t_P$ according to the time-dependent Schr\"odinger equation. If $t_P$ is sufficiently large, the probability that the electron is subsequently found in the eigenstate $\psi_{R,\beta}$ of $\mathcal{H}_R$ at $t>t_P$, is, to first order and per unit $t_P$,
\begin{equation} \label{eq_probability_rate}
P_{\alpha \beta} = \frac{2\pi}{\hbar} \left| \langle \psi_{R,\beta} | \mathcal{H}-\mathcal{H}_L | \psi_{L,\alpha} \rangle \right|^2 \delta \left( E_{L,\alpha} - E_{R,\beta} \right) \, ,
\end{equation}
where $E_{L,\alpha}$ and $E_{R,\beta}$ are the eigenvalues corresponding to the initial state $\psi_{L,\alpha}$ and final state $\psi_{R,\beta}$, respectively, and $\delta$ is Dirac's delta function \cite{oppenheimer1928three}. The conservation of energy is related to the perturbation being constant in time. In (\ref{eq_probability_rate}), it is assumed that each set of eigenfunctions $\psi_{L,\alpha}$ and $\psi_{R,\beta}$ is discrete and orthonormal. We consider a rectangular domain with finite sides $L_x$ and $L_y$ in the plane parallel to the junction and prescribe the periodic boundary conditions $\psi_{L,\alpha/R,\beta}(x=0,y,z)=\psi_{L,\alpha/R,\beta}(x=L_x,y,z)$, $\psi_{L,\alpha/R,\beta}(x,y=0,z)=\psi_{L,\alpha/R,\beta}(x,y=L_y,z)$. In addition, we consider a finite length $L_z$ of the metal region in the $z$ direction with the hard-wall boundary condition $\psi_{L,\alpha}(x,y,z=-L_z)=0$. This way, both energy spectra are discrete and the corresponding eigenfunctions normalizable. Later, we will take the limit as $L_x,L_y,L_z$ go to infinite in order to recover the continuous case. It should be noted that, contrary to the standard time-dependent perturbation theory \cite{rudan2015physics_S}, the matrix element in (\ref{eq_probability_rate}) must be computed between eigenstates of different Hamiltonians. Also, the perturbation Hamiltonian must be evaluated with respect to the initial Hamiltonian. The validity of (\ref{eq_probability_rate}) rests on the assumption that the two sets of eigenstates of $\mathcal{H}_L$ and $\mathcal{H}_L$ are ``almost orthogonal'' to each other, in particular that $\langle \psi_{R,\beta} | \psi_{L,\alpha} \rangle \ll 1$ \cite{oppenheimer1928three}. Similary, one has for the probability rate of the inverse transition
\begin{equation} \label{eq_probability_rate_inverse}
P_{\beta \alpha} = \frac{2\pi}{\hbar} \left| \langle \psi_{L,\alpha} | \mathcal{H}-\mathcal{H}_R | \psi_{R,\beta} \rangle \right|^2 \delta \left( E_{L,\alpha} - E_{R,\beta} \right) = P_{\alpha \beta} \, ,
\end{equation}
where the last equality follows from the delta function and the Hermiticity of the various Hamiltonians.

From (\ref{eq_potential})-(\ref{eq_potential_right}) we get
\begin{equation} \label{eq_separability}
\mathcal{H} = \left\{
\begin{array}{ll}
\mathcal{H}_L, & \quad z<0 \\
\mathcal{H}_R, & \quad z>0
\end{array}
\right.
\end{equation}
which means that $\mathcal{H}$ satisfies the separability property of Bardeen's model Hamiltonian, i.e., that $\mathcal{H} - \mathcal{H}_L \ne 0$ only in regions of space where $\mathcal{H} - \mathcal{H}_R \equiv 0$ \cite{bardeen1961tunnelling_S} \footnote{Bardeen's theory is often introduced by writing the Hamiltonian in the form $\mathcal{H} = \mathcal{H}_L + \mathcal{H}_R + \mathcal{H}_T$, with $\mathcal{H}_T$ being the ``transfer'' Hamiltonian, despite the fact that Bardeen himself did not make use of such decomposition in its original paper \cite{bardeen1961tunnelling_S}. In our case, we get from (\ref{eq_separability})
\begin{equation}
\mathcal{H}_T = \left\{
\begin{array}{ll}
-\mathcal{H}_R, & \quad z<0 \\
-\mathcal{H}_L, & \quad z>0
\end{array}
\right.
\end{equation}
but this Hamiltonian does not correspond to either of the perturbation Hamiltonians that appear in (\ref{eq_probability_rate}) or (\ref{eq_probability_rate_inverse}). See also the discussion in \citep{duke1969tunneling_S}.}. Since $\mathcal{H} - \mathcal{H}_L \equiv 0$ for $z<0$, the matrix element in (\ref{eq_probability_rate}) can be written as
\begin{align}
\langle \psi_{R,\beta} | \mathcal{H}-\mathcal{H}_L | \psi_{L,\alpha} \rangle &= \int_0^{L_x} dx \int_0^{L_y} dy \int_{-\infty}^{\infty} dz \, \psi_{R,\beta}^* \left( \mathcal{H}-\mathcal{H}_L \right) \psi_{L,\alpha} \nonumber \\
&= \int_{\Omega_R} \psi_{R,\beta}^* \left( \mathcal{H}-\mathcal{H}_L \right) \psi_{L,\alpha} d^3 r \, ,
\end{align}
where $\mathbf{r}=(x,y,z)$ and $\Omega_R = \{\mathbf{r};0<x<L_x,0<y<L_y,0<z<\infty\}$. Noting that $\mathcal{H} - \mathcal{H}_R \equiv 0$ for $z>0$, we can get the symmetric expression
\begin{align}
\langle \psi_{R,\beta} | \mathcal{H}-\mathcal{H}_L | \psi_{L,\alpha} \rangle &= \int_{\Omega_R} \left[ \psi_{R,\beta}^* \left( \mathcal{H}-\mathcal{H}_L \right) \psi_{L,\alpha} -  \psi_{L,\alpha} \left( \mathcal{H}-\mathcal{H}_R \right) \psi_{R,\beta}^* \right] d^3 r \nonumber \\
&= \int_{\Omega_R} \left[ \psi_{R,\beta}^* \left( \mathcal{T}-E_{L,\alpha} \right) \psi_{L,\alpha} -  \psi_{L,\alpha} \left( \mathcal{T}-E_{R,\beta} \right) \psi_{R,\beta}^* \right] d^3 r \, .
\end{align}
Due to the delta function in (\ref{eq_probability_rate}), we are only interested in the case $E_{L,\alpha} = E_{R,\beta}$, for which
\begin{equation}
\langle \psi_{R,\beta} | \mathcal{H}-\mathcal{H}_L | \psi_{L,\alpha} \rangle = \int_{\Omega_R} \left[ \psi_{R,\beta}^* \mathcal{T} \psi_{L,\alpha} -  \psi_{L,\alpha} \mathcal{T} \psi_{R,\beta}^* \right] d^3 r \, .
\end{equation}
Applying Green's theorem, we finally get
\begin{equation} \label{eq_matrix_element}
\langle \psi_{R,\beta} | \mathcal{H}-\mathcal{H}_L | \psi_{L,\alpha} \rangle = -\mathrm{i} \hbar \int_{\Sigma_R} \mathbf{n} \cdot \mathbf{J}_{\beta\alpha} d^2 r \, ,
\end{equation}
where $\Sigma_R$ is the surface of $\Omega_R$, $\mathbf{n}$ is the unit vector normal to $\Sigma_R$ pointing in the outward direction, and
\begin{equation} \label{eq_current_density}
\mathbf{J}_{\beta\alpha} = -\frac{\mathrm{i}\hbar}{2} \hat{m}^{-1} \left[ \psi_{R,\beta}^* \nabla \psi_{L,\alpha} -  \psi_{L,\alpha} \nabla \psi_{R,\beta}^* \right]
\end{equation}
is the matrix element of the probability current density operator between the states $\psi_{L,\alpha}$ and $\psi_{R,\beta}$.

Let us now start to pick up all the ingredients that we need to calculate (\ref{eq_matrix_element}). The eigenfunctions and corresponding eigenvalues of $\mathcal{H}_L$ are \cite{rudan2015physics_S}
\begin{align}
\psi_{L,\alpha}(\mathbf{r}) &\equiv \psi_L(\mathbf{k}_L;\mathbf{r}) = \frac{1}{\sqrt{L_x L_y}} e^{\mathrm{i} (k_{xL} x + k_{yL} y)} b \times
\left\{
\begin{array}{ll}
0, & \quad z<-L_z \\
\cos(k_z z + \varphi), & \quad -L_z<z<0 \\
\cos(\varphi)e^{-\kappa z}, & \quad z>0 
\end{array}
\right. \label{eq_wavefunction_left} \\
E_{L,\alpha} &\equiv E_L(\mathbf{k}_L) = \frac{\hbar^2}{2}\left( \frac{k_{xL}^2}{m_x} + \frac{k_{yL}^2}{m_y} \right) + E_z \, , \\
E_z &= V_0 + \frac{\hbar^2 k_z^2}{2 m_z} \, , \\
\kappa &= \frac{\sqrt{- 2 m_z E_z}}{\hbar} \, , \\
\varphi &= \arctan(\kappa/k_z) \, ,
\end{align}
where $\mathbf{k}_L = (k_{xL},k_{yL},k_z)$ and it is assumed that $E_z < 0$ \footnote{We limit the discussion to the case $E_z < 0$ because (\ref{eq_probability_rate}) looses validity if $\langle \psi_{R,\beta} | \psi_{L,\alpha} \rangle$ is not a small number.}.
Because of the periodic boundary conditions, the transverse components of the wavevector are quantized as $k_{xL} = 2 \pi l/L_x$, $k_{yL} = 2 \pi m/L_y$ ($l,m$ integers). As for $k_z$, the allowed values are the roots of the transcendental equations $k_z L_z - \pi (n-1/2) = \varphi$ in the interval $0 < k_z <\sqrt{-2 m_z V_0}/\hbar$. For large $L_z$, we have $k_z \approx \pi n/L_z$ ($n$ positive integer). The constant $b$ can be obtained from the normalization condition
\begin{equation}
1 = |b|^2 \left[ \int_{-L_z}^{0} \cos^2(k_z z + \varphi) dz + \cos^2(\varphi) \int_{0}^{\infty} e^{-2 \kappa z} dz \right] = \frac{|b|^2}{2} \left( L_z+\frac{1}{\kappa}\right) \approx |b|^2 \frac{L_z}{2} \, ,
\end{equation}
where only the leading term in $L_z$ has been kept. Thus, up to an unimportant phase, $b = \sqrt{2/L_z}$.

The solutions of the eigenvalue problem of $\mathcal{H}_R$ are \cite{miller2008quantum_S}
\begin{align}
\psi_{R,\beta}(\mathbf{r}) &\equiv \psi_{R,i}(\mathbf{k}_{\parallel R}; \mathbf{r}) = \frac{1}{\sqrt{L_x L_y}} e^{\mathrm{i} (k_{xR} x + k_{yR} y)} c \times
\left\{
\begin{array}{ll}
\mathrm{Ai}(\zeta), & \quad z<a \\
0, & \quad z>a 
\end{array}
\right. \label{eq_wavefunction_right} \\
E_{R,\beta} &\equiv E_{R,i}(\mathbf{k}_{\parallel R}) = \frac{\hbar^2}{2}\left( \frac{k_{xR}^2}{m_x} + \frac{k_{yR}^2}{m_y} \right) + E_i \, , \\
E_i &= - qF \left( a - \frac{| \zeta_i |}{k_F} \right) , \\
\zeta &= -k_F \left( z+\frac{E_i}{qF} \right), \\
k_F &= \left( \frac{2 m_z qF}{\hbar^2} \right)^{1/3} ,
\end{align}
where $\mathbf{k}_{\parallel R} = (k_{xR},k_{yR})$, $\mathrm{Ai}$ is Airy's function and $\zeta_i$ are its zeros, which can be approximated as ($i$ positive integer) \cite{abramowitz1964handbook_S}
\begin{equation} \label{eq_zeros_S}
\zeta_i \approx - \left[ \frac{3 \pi}{8} (4 i-1) \right]^{2/3} .
\end{equation}
The constant $c$ can be obtained from the normalization condition
\begin{equation}
1 = |c|^2 \int_{-\infty}^{a} \mathrm{Ai}^2(\zeta) dz = \frac{|c|^2}{k_F} \int_{\zeta_i}^{\infty} \mathrm{Ai}^2(\zeta) d\zeta \, ,
\end{equation}
where the last integral can be evaluated using integration by parts and the fact that $\mathrm{Ai}$ is a solution of Airy's equation $\mathrm{Ai}'' = \zeta \mathrm{Ai}$ (the prime symbol indicates derivative with respect to $\zeta$):
\begin{equation}
\int_{\zeta_i}^{\infty} \mathrm{Ai}^2(\zeta) d\zeta = - \int_{\zeta_i}^{\infty} 2 \mathrm{Ai}(\zeta) \mathrm{Ai}'(\zeta) \zeta d\zeta = - \int_{\zeta_i}^{\infty} 2 \mathrm{Ai}'(\zeta) \mathrm{Ai}''(\zeta) d\zeta = \mathrm{Ai}'^2(\zeta_i) \, .
\end{equation}
Therefore, $c = \sqrt{k_F}/|\mathrm{Ai}'(\zeta_i)|$, in which we can use the approximated expression \cite{abramowitz1964handbook_S}
\begin{equation} \label{eq_derivative_at_zeros_S}
|\mathrm{Ai}'(\zeta_i)| \approx \frac{1}{\sqrt{\pi}} \left[ \frac{3 \pi}{8} (4 i-1) \right]^{1/6} .
\end{equation}

The surface $\Sigma_R$ in (\ref{eq_matrix_element}) is made up of six faces. By inserting (\ref{eq_wavefunction_left}) and (\ref{eq_wavefunction_right}) into (\ref{eq_matrix_element}), it can be shown that the integrals over the two faces at $x=0$ and $x=L_x$, as well as the integrals over the two faces at $y=0$ and $y=L_y$, cancel out each other exactly \footnote{For example, $J_{x,\beta \alpha} (x=L_x,y,z)- J_{x,\beta \alpha} (x=0,y,z) \propto e^{\mathrm{i} (k_{xL} -k_{xR}) L_x} - 1=0$ because of the periodic boundary conditions.}. The integral over the face at $z=\infty$ is also zero because $\psi_{L,\alpha}$ is vanishingly small. We are only left with the integral over the face at $z=0$:
\begin{align}
\langle \psi_{R,\beta} | \mathcal{H}-\mathcal{H}_L | \psi_{L,\alpha} \rangle &= \frac{\hbar^2}{2 m_z} \sqrt{\frac{2}{L_z}} \frac{\sqrt{k_F}}{|\mathrm{Ai}'(\zeta_i)|} \cos(\varphi) \left[ \mathrm{Ai}(\zeta) \frac{d}{dz} e^{-\kappa z} - e^{-\kappa z} \frac{d}{dz} \mathrm{Ai}(\zeta) \right]_{z=0} \nonumber \\
& \quad \times \frac{1}{L_x} \int_0^{L_x} e^{\mathrm{i} (k_{xL} -k_{xR}) x} dx \frac{1}{L_y} \int_0^{L_y} e^{\mathrm{i} (k_{yL} -k_{yR}) y} dy \nonumber \\
&= - \frac{\hbar^2}{2 m_z} \sqrt{\frac{2}{L_z}} \frac{\sqrt{k_F}}{|\mathrm{Ai}'(\zeta_i)|} \cos(\varphi) \left[ \kappa \mathrm{Ai}(\zeta_0) - k_F \mathrm{Ai'}(\zeta_0) \right] \delta_{k_{xL},k_{xR}} \delta_{k_{yL},k_{yR}} \, ,\label{eq_matrix_element_2}
\end{align}
where
\begin{equation}
\zeta_0 \equiv \zeta(z=0) = - k_F \frac{E_i}{qF} = k_F a - |\zeta_i|
\end{equation}
and $\delta$ is Kronecker's delta function. Conservation of transverse momentum is a consequence of the translational symmetry along $x$ and $y$. Combined with energy conservation, it implies that $E_z = E_i$ and thus $\zeta_0 = ( \kappa/ k_F )^2$. Assuming $\zeta_0 \gg 1$ (which is consistent with $\langle \psi_{R,\beta} | \psi_{L,\alpha} \rangle \ll 1$), we can substitute in (\ref{eq_matrix_element_2}) the asymptotic expressions
\begin{align}
\mathrm{Ai}(\zeta_0) &\approx \frac{e^{-\gamma_0}}{2 \sqrt{\pi} \zeta_0^{1/4}} \, ,\\
\mathrm{Ai}'(\zeta_0) &\approx - \frac{\zeta_0^{1/4} e^{-\gamma_0}}{2 \sqrt{\pi}} \, ,
\end{align}
where $\gamma_0 = (2/3) \zeta_0^{3/2}$ \cite{abramowitz1964handbook_S}, to get
\begin{align}
\langle \psi_{R,\beta} | \mathcal{H}-\mathcal{H}_L | \psi_{L,\alpha} \rangle &= - \frac{\hbar^2}{2 m_z} \sqrt{\frac{2}{L_z}} \frac{\sqrt{k_F}}{|\mathrm{Ai}'(\zeta_i)|} \cos(\varphi) \left[ \kappa + k_F \sqrt{\zeta_0} \right] \frac{e^{-\gamma_0}}{2 \sqrt{\pi} \zeta_0^{1/4}} \delta_{k_{xL},k_{xR}} \delta_{k_{yL},k_{yR}} \nonumber \\
&= - \frac{\hbar^2}{2 m_z} \sqrt{\frac{2}{L_z}} \frac{k_F}{|\mathrm{Ai}'(\zeta_i)|} \frac{2 k_z \sqrt{\kappa}}{\sqrt{k_z^2+\kappa^2}} \frac{e^{-\gamma_0}}{2 \sqrt{\pi}} \delta_{k_{xL},k_{xR}} \delta_{k_{yL},k_{yR}} \, . \label{eq_matrix_element_3}
\end{align}
Finally, plugging (\ref{eq_matrix_element_3}) into (\ref{eq_probability_rate}), we obtain
\begin{align}
P_{\alpha \beta} = \frac{1}{h} \left( \frac{\hbar^2}{2 m_z} \right)^{2} \frac{2 \pi}{L_z} \frac{k_F^2}{\mathrm{Ai}'^2(\zeta_i)} \frac{4 k_z^2 \kappa}{k_z^2+\kappa^2} e^{-2\gamma_0} \delta_{k_{xL},k_{xR}} \delta_{k_{yL},k_{yR}} \delta \left( E_{L,\alpha} - E_{R,\beta} \right) \, . \label{eq_probability_rate_2}
\end{align}

\section{Model of longitudinal diffusion} \label{sec_model_diffusion}

Suppose that the states in the metal ($L$) are populated according to a Fermi-Dirac distribution with Fermi level $\mu_L$:
\begin{equation} \label{eq_fermi}
f_{L}(E_{L,\alpha}) = \frac{1}{\exp\left( \frac{E_{L,\alpha}-\mu_{L} }{k_B T} \right) + 1 }
\end{equation}
with $k_B$ Boltzmann's constant and $T$ the temperature. As for the semiconductor, we cannot assume that the states are in equilibrium because a current has to flow in the $x$ direction as shown in Fig.~1 of the main text. In order to compute the population of such states, we assume semiclassical diffusive transport and make use of Boltzmann's transport equation \cite{rudan2015physics_S}. 

Let $f_i(x,\mathbf{k}_{\parallel R})$ be the distribution function in the four-dimensional phase space associated with the $i$-th subband \footnote{$f_i$ is independent of $y$ because of the translational symmetry along $y$.}. Under the assumption that the electric potential is uniform along $x$, Boltzmann's equation reads
\begin{equation} \label{eq_bte}
v_x \frac{\partial f_i}{\partial x} = C, \quad 0<x<L_x
\end{equation}
where $v_x = \hbar k_{xR}/m_x$ is the longitudinal carrier velocity and the transitions from the metal to the semiconductor and vice versa due to vertical tunneling are included through a collision term $C$ \footnote{Other types of scattering, which could be responsible for a finite carrier mobility in the semiconductor, are here neglected.}:
\begin{align} \label{eq_collision}
C &= \sum_{\alpha} f_L(E_{L,\alpha}) P_{\alpha \beta} ( 1-f_i) - f_i P_{\beta \alpha} \left[ 1 - f_L(E_{L,\alpha}) \right] \nonumber \\
&= \sum_{\alpha} P_{\alpha \beta} \left[ f_L(E_{L,\alpha}) - f_i \right] \, .
\end{align}
Note that expression (\ref{eq_collision}) takes into account Pauli's exclusion principle. Using (\ref{eq_probability_rate_2}), we get
\begin{equation} \label{eq_collision_2}
C = \left[ f_L(E_{R,\beta}) - f_i \right] \sum_{\alpha} P_{\alpha \beta} = \frac{ f_L(E_{R,\beta}) - f_i }{\tau_i} \, ,
\end{equation}
where the relaxation time $\tau_i$ is defined as 
\begin{equation}
\frac{1}{\tau_i} = \sum_{\alpha} P_{\alpha \beta} = \sum_{k_z} \frac{1}{h} \left( \frac{\hbar^2}{2 m_z} \right)^{2} \frac{2 \pi}{L_z} \frac{k_F^2}{\mathrm{Ai}'^2(\zeta_i)} \frac{4 k_z^2 \kappa}{k_z^2+\kappa^2} e^{-2\gamma_0} \delta \left( E_z - E_i \right) \, .
\end{equation}
Going to the limit of large $L_z$, we can replace
\begin{equation}
\sum_{k_z} \rightarrow \int \frac{L_z}{\pi} d k_z
\end{equation}
so that
\begin{equation}
\frac{1}{\tau_i} = \int_{0}^{\frac{\sqrt{-2 m_z V_0}}{\hbar}} d k_z \frac{1}{h} \left( \frac{\hbar^2}{2 m_z} \right)^{2} 2 \frac{k_F^2}{\mathrm{Ai}'^2(\zeta_i)} \frac{4 k_z^2 \kappa}{k_z^2+\kappa^2} e^{-2\gamma_0} \delta \left( E_z - E_i \right)
\end{equation}
and, with the change of variables $k_z \rightarrow E_z$,
\begin{align} \label{eq_relaxation_time_S}
\frac{1}{\tau_i} &= \int_{V_0}^{0} d E_z \frac{1}{h} \frac{\hbar^2}{2 m_z} \frac{k_F^2}{\mathrm{Ai}'^2(\zeta_i)} \frac{4 \sqrt{-E_z (E_z - V_0)}}{-V_0} e^{-2\gamma_0} \delta \left( E_z - E_i \right) \nonumber \\
&= \left\{
\begin{array}{ll}
\frac{1}{h} \frac{\hbar^2}{2 m_z} \frac{k_F^2}{\mathrm{Ai}'^2(\zeta_i)} \frac{4 \sqrt{-E_i (E_i - V_0)}}{-V_0} e^{-2\gamma_0}, & \quad V_0 < E_i < 0 \\
0, & \quad \text{otherwise}
\end{array}
\right.
\end{align}
Besides the subband index $i$, the relaxation time depends on the parameters $m_z$, $F$, $a$, and $V_0$. It can be shown that, replacing the eigenfunctions (\ref{eq_wavefunction_left}) of $\mathcal{H}_L$ by their WKB \cite{messiah1961quantum} approximation
\begin{equation}
\psi_{L,\alpha}^{\mathrm{WKB}}(\mathbf{r}) = \frac{1}{\sqrt{L_x L_y}} e^{\mathrm{i} (k_{xL} x + k_{yL} y)} \sqrt{\frac{2}{L_z}} \times
\left\{
\begin{array}{ll}
0, & \quad z<-L_z \\
\cos\left(k_z z + \frac{\pi}{4}\right), & \quad -L_z<z<0 \\
\frac{1}{2} \sqrt{\frac{k_z}{\kappa}} e^{-\kappa z}, & \quad z>0 
\end{array}
\right.
\end{equation}
the expression of $\tau_i$ simplifies to
\begin{equation} \label{eq_relaxation_time_wkb}
\frac{1}{\tau_i^{\mathrm{WKB}}} = \left\{
\begin{array}{ll}
\frac{1}{h} \frac{\hbar^2}{2 m_z} \frac{k_F^2}{\mathrm{Ai}'^2(\zeta_i)} e^{-2\gamma_0}, & \quad V_0 < E_i < 0 \\
0, & \quad \text{otherwise}
\end{array}
\right.
\end{equation}
where $V_0$ appears only as an energy cut-off. This last formulation, which does not depend on the precise bandstructure of the metal, could be useful for treating injection from the metal to the valence band of the semiconductor.

\begin{figure}[t]
\centering
\scalebox{0.55}[0.55]{\includegraphics*[viewport=200 290 530 420]{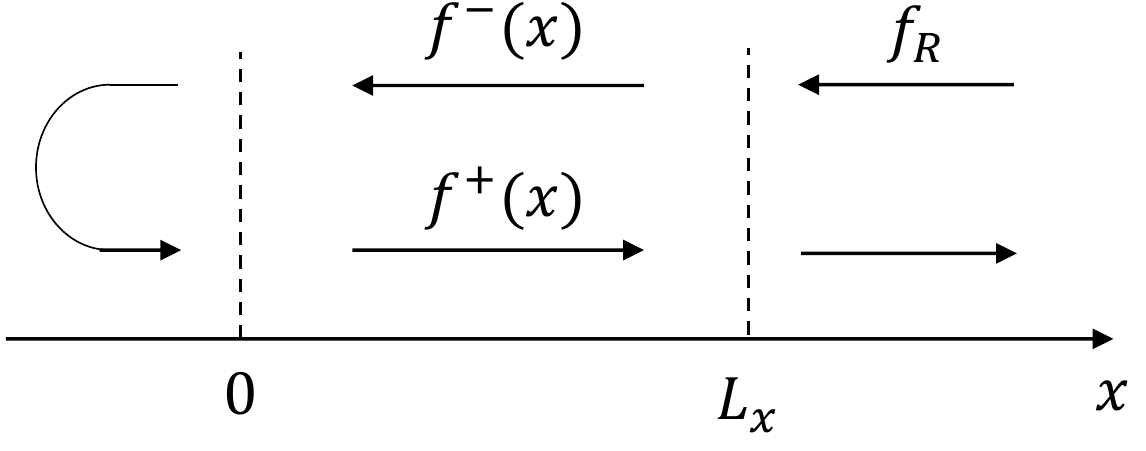}}
\caption{Schematic description of the current fluxes and the boundary conditions of Boltzmann's equation.
}
\label{fig_boltzmann}
\end{figure}
Let $f_i^+$ and $f_i^-$ denote the distribution functions of right-going and left-going states, respectively, i.e., $f_i^{\pm}(x,k_{xR},k_{yR}) = f_i(x,\pm k_{xR},k_{yR})$ with $k_{xR}>0$. We impose the boundary conditions
\begin{align}
f_i^-(x = L_x,\mathbf{k}_{\parallel R}) &= f_R(E_{R,\beta}) \, , \label{eq_bcm} \\
f_i^+(x = 0,\mathbf{k}_{\parallel R}) &= f_i^-(x = 0,\mathbf{k}_{\parallel R}) \, , \label{eq_bcp}
\end{align}
where $f_R$ is a Fermi-Dirac function similar to (\ref{eq_fermi}) with $\mu_L$ replaced by $\mu_R$ (see Fig.~\ref{fig_boltzmann}). The latter condition makes sure that the longitudinal current vanishes at $x=0$. The Boltzmann equations (\ref{eq_bte}) for different wavevectors are independent of each other expect for $\pm k_{xR}$. The solutions are
\begin{align}
f_i^{\pm} &= -\left[ f_L(E_{R,\beta}) - f_R(E_{R,\beta}) \right] e^{- \frac{L_x \pm x}{|v_x| \tau_i}} + f_L(E_{R,\beta}) \, .
\end{align}

The current (per unit width) from the semiconductor to the metal can be obtained by summing the net flux $|v_x|\left(f_i^+-f_i^-\right)$ at $x = L_x$ over all semiconductor states per unit area, multiplying by $2$ for spin degeneracy and multiplying by the electronic charge $q$:
\begin{align}
I &= \frac{2q}{L_x L_y} \sum_{k_{xR}>0} \sum_{k_{yR}} \sum_{i} |v_x| \left(f_i^+-f_i^-\right)_{x=L_x} \nonumber \\
&= \frac{2 q}{L_x L_y} \sum_{k_{xR}>0} \sum_{k_{yR}} \sum_{i} |v_x| \left( 1 - e^{-\frac{2 L_x}{|v_x| \tau_i}}\right) \left[ f_L(E_{R,\beta}) - f_R(E_{R,\beta}) \right] .
\end{align}
Going to the limit of large $L_x,L_y$, we can replace
\begin{equation}
\sum_{k_{xR}} \rightarrow \int \frac{L_x}{2\pi} d k_{xR} \, , \quad \sum_{k_{yR}} \rightarrow \int \frac{L_y}{2\pi} d k_{yR}
\end{equation}
to get
\begin{equation} \label{eq_current_double_integral}
I = \frac{2q}{( 2 \pi)^2} \int_{0}^{\infty} d k_{xR} \int_{-\infty}^{\infty} d k_{yR} \sum_{i} |v_x| \left( 1 - e^{-\frac{2 L_x}{|v_x| \tau_i}}\right) \left[ f_L(E_{R,\beta}) - f_R(E_{R,\beta}) \right] .
\end{equation}
Finally, with the change of variables $k_{xR} \rightarrow \varepsilon = \hbar^2 k_{xR}^2 / (2 m_x) + E_i$, we obtain the Landauer formula \cite{datta1997electronic_S}
\begin{align} \label{eq_current}
I &= \frac{2q}{h} \sum_{i} \int_{E_i}^{\infty} d \varepsilon \, \left( 1 - e^{-\frac{2 L_x}{|v_x| \tau_i}}\right) \frac{1}{2 \pi} \int_{-\infty}^{\infty} d k_{yR} \left[ f_L\left( \frac{\hbar^2 k_{yR}^2}{2 m_y} + \varepsilon \right) - f_R\left( \frac{\hbar^2 k_{yR}^2}{2 m_y} + \varepsilon \right) \right] \nonumber \\
&= \frac{2q}{h} \sum_{i} \int_{E_i}^{\infty} d \varepsilon \, \left( 1 - e^{-\frac{2 L_x}{|v_x| \tau_i}}\right) \left[ F_L(\varepsilon) - F_R(\varepsilon) \right] \nonumber \\
&=  \frac{2q}{h} \int_{-\infty}^{\infty} d \varepsilon \, \overline{T}(\varepsilon) \left[ F_L(\varepsilon) - F_R(\varepsilon) \right] ,
\end{align}
where the longitudinal velocity must be computed as $|v_x| = \sqrt{ 2 (\varepsilon-E_i) / m_x}$, the supply function $F_{L/R}$ is defined as
\begin{equation} \label{eq_supply}
F_{L/R}(\varepsilon) = \sqrt{\frac{m_y k_B T}{2 \pi \hbar^2}} \mathcal{F}_{-1/2} \left( \frac{\mu_{L/R} - \varepsilon}{k_B T} \right)
\end{equation}
with $\mathcal{F}_{-1/2}$ the Fermi-Dirac integral of order $-\tfrac{1}{2}$ \cite{rudan2015physics_S}, $\overline{T}$ is the transmission function
\begin{equation} \label{eq_transmission_function_S}
\overline{T}(\varepsilon) = \sum_i T_i(\varepsilon) \, ,
\end{equation}
with the transmission probability $T_i$ given by
\begin{equation} \label{eq_transmission_S}
T_i(\varepsilon) = \left\{
\begin{array}{ll}
0, & \quad \varepsilon < E_i \\
1 - e^{-\frac{2 L_x}{\lambda_i}}, & \quad \varepsilon > E_i 
\end{array}
\right.
\end{equation}
and $\lambda_i = |v_x| \tau_i$. For $\varepsilon > E_i$, we can have the asymptotic behaviors 
\begin{align}
L_x \gg \lambda_i &\text{:} \quad T_i(\varepsilon) \approx 1 \label{eq_transmission_gg} \\
L_x \ll \lambda_i &\text{:} \quad T_i(\varepsilon) \approx \frac{2 L_x}{\lambda_i} \label{eq_transmission_ll}
\end{align}
Approximation (\ref{eq_transmission_gg}) holds, in particular, as $\varepsilon \rightarrow E_i^+$ (resonant tunneling). Note also that, when (\ref{eq_transmission_ll}) is satisfied, $T_i$ decays as a function of energy as $1/\sqrt{\varepsilon}$.

The low bias conductance per unit width $G$ can be evaluated from (\ref{eq_current}). Let $\mu_L=\mu$ and $\mu_R=\mu-qV_D$. We have
\begin{equation} \label{eq_cond_S}
G = \left.\frac{\partial I}{\partial V_D}\right|_{V_D=0} = \frac{2q^2}{h} \int_{-\infty}^{\infty} d \varepsilon \, \overline{T}(\varepsilon) \left(-\frac{\partial F_0}{\partial \varepsilon}\right) ,
\end{equation}
where $F_0(\varepsilon)$ is given by (\ref{eq_supply}) with $\mu_{L/R}$ replaced by $\mu$.

Assume for simplicity that only the first subband contributes to transport. In the two limiting cases when either (\ref{eq_transmission_gg}) or (\ref{eq_transmission_ll}) are satisfied over the whole energy range of interest for transport (from $E_1$ to few $k_B T$'s above $\max\{E_1,\mu\}$), it is possible to derive analytical expressions for $G$. If $L_x \gg \lambda_1$ in this energy range, it follows immediately from (\ref{eq_cond_S})
\begin{equation} \label{eq_cond_approx1_S}
G = \frac{2q^2}{h} F_0(E_1) \, .
\end{equation}
To work out the expression of $G$ in the other limiting case when $L_x \ll \lambda_1$ in most of the energy window for transport\footnote{The inequality does not hold for energies close to $E_1$ but their contribution becomes increasingly smaller as $\tau_1$ increases.}, it is convenient to go back to the double-integral formulation of the tunneling current in (\ref{eq_current_double_integral}) and do the change of variables $k_{xR} \rightarrow E_{R,\beta}$:
\begin{equation}
I = \frac{2q}{h} \int_{E_1}^{\infty} d E_{R,\beta} \left( \frac{1}{\pi} \int_{0}^{k_{y\mathrm{max}}} d k_{yR} \frac{2 L_x}{|v_x| \tau_1} \right) \left[ f_L(E_{R,\beta}) - f_R(E_{R,\beta}) \right] \, ,
\end{equation}
where $k_{y\mathrm{max}} = \sqrt{2 m_y (E_{R,\beta}-E_1)}/\hbar$ and $|v_x| = \sqrt{ 2 [E_{R,\beta}-\hbar^2 k_{yR}^2/(2 m_y)-E_1] / m_x}$. The integral over the transverse wavevector can be easily computed with the change of variables $k_{yR} \rightarrow \arcsin(k_{yR}/k_{y\mathrm{max}})$ to give
\begin{align}
I &= \frac{2q}{h} \int_{E_1}^{\infty} d E_{R,\beta} \frac{k_{y\mathrm{max}} L_x}{\sqrt{2 (E_{R,\beta}-E_1)/m_x} \tau_1} \left[ f_L(E_{R,\beta}) - f_R(E_{R,\beta}) \right] \nonumber \\
&= \frac{2q}{h} \frac{\sqrt{m_x m_y}}{\hbar} \frac{L_x}{\tau_1} \int_{E_1}^{\infty} d E_{R,\beta} \left[ f_L(E_{R,\beta}) - f_R(E_{R,\beta}) \right] \, .
\end{align}
Letting $\mu_L=\mu$ and $\mu_R=\mu-qV_D$, we finally get
\begin{equation} \label{eq_cond_approx2_S}
G = \left.\frac{\partial I}{\partial V_D}\right|_{V_D=0} = \frac{2q^2}{h} \frac{\sqrt{m_x m_y}}{\hbar} \frac{L_x}{\tau_1} \int_{E_1}^{\infty} d E_{R,\beta} \left(-\frac{\partial f_0}{\partial E_{R,\beta}}\right) = \frac{2q^2}{h} \frac{\sqrt{m_x m_y}}{\hbar} \frac{L_x}{\tau_1} f_0(E_1) \, ,
\end{equation}
where $f_0$ is a Fermi-Dirac function similar to (\ref{eq_fermi}) with $\mu_L$ replaced by $\mu$.

\section{Green function algorithm}

\begin{figure}[t]
\centering
\scalebox{0.55}[0.55]{\includegraphics*[viewport=140 180 820 420]{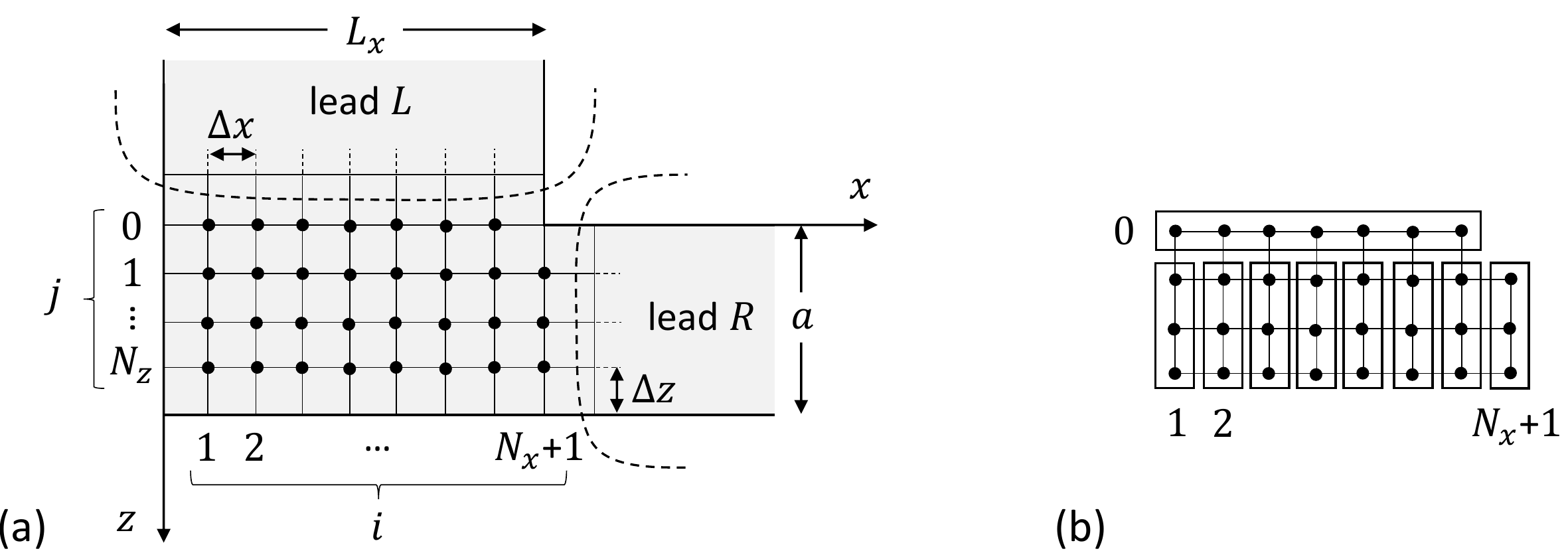}}
\caption{(a) Simulation domain and rectangular grid of the Green function calculation. The discretization steps are $\Delta x = 0.3$~nm and $\Delta z = 0.2$~nm. The black dots indicate the grid nodes that compose the ``channel'' region. (b) Partitioning of the channel region into layers for the Green function algorithm.}
\label{fig_negf}
\end{figure}
We discretize the Hamiltonian in (\ref{eq_hamiltonian}) using finite differences on a two-dimensional rectangular grid (Fig.~\ref{fig_negf}a). The same linear potential profile as in (\ref{eq_potential}) is assumed. The transmission function is computed as \cite{datta1997electronic_S}
\begin{equation} \label{eq_transm1}
\overline{T}(\varepsilon) = \mathrm{tr} \left[ \Gamma^L G^r \Gamma^R G^a \right] ,
\end{equation}
where the symbol $\mathrm{tr}$ indicates the trace, $G^r$ is the retarded Green function, $G^a=G^{r\dagger}$, $\Gamma^{L/R} = \mathrm{i} (\Sigma^{r,L/R}-{\Sigma^{a,L/R}})$, with $\Sigma^{r,L/R}$ the retarded self-energy representing the renormalization of the Hamiltonian of the channel region (black dots in Fig.~\ref{fig_negf}a) due to the presence of the semi-infinite left/right lead, and $\Sigma^{a,L/R} = ( \Sigma^{r,L/R} )^\dagger$. The channel region is partitioned into layers as shown in Fig.~\ref{fig_negf}b. Using matrix block notation and noting that the only non-null block of $\Sigma^{r,L}$ is $\Sigma^{r,L}_{0,0}$ and the only non-null block of $\Sigma^{r,R}$ is $\Sigma^{r,R}_{N_x+1,N_x+1}$, (\ref{eq_transm1}) can be rewritten as
\begin{equation}
\overline{T}(\varepsilon) = \mathrm{tr} \left[ \Gamma^L_{0,0} G^r_{0,N_x+1} \Gamma^R_{N_x+1,N_x+1} {G^a_{N_x+1,0}} \right] . \label{eq_transm2}
\end{equation}
The self-energy of the left lead is computed analytically using the prescription given in \cite{datta1997electronic_S}:
\begin{align}
\Sigma^{r,L}_{0,0}(i,i') &= \sum_{m=1}^{N_x} \chi_m(i) \sigma_m \chi_m(i') \, , \\
\chi_m(i) &= \sqrt{\frac{2}{N_x+1}} \sin(k_x i) \, , \\
\sigma_m &= t_z \times
\left\{
\begin{array}{ll}
\lambda-1+\sqrt{\lambda^2-2\lambda}, & \quad \lambda < 0 \\
\lambda-1-\sqrt{\lambda^2-2\lambda}, & \quad \lambda > 2 \\
\lambda-1-\mathrm{i}\sqrt{2\lambda-\lambda^2}, & \quad 0 < \lambda < 2 
\end{array}
\right. \\
\lambda &= \frac{\varepsilon - V_0 - 2 t_x (1 - \cos k_x)}{2 t_z} \, , \\
k_x &= \frac{\pi m}{N_x+1} \, ,
\end{align}
where $t_x = \hbar^2/(2 m_x \Delta_x^2)$ and similarly for $t_z$. The self-energy of the right lead is obtained numerically using a well-known iterative algorithm \cite{sancho1985highly}. The matrix block $G^r_{0,N_x+1}$ is computed through a combination of the recursive and decimation algorithms \cite{low2009electronic}, modified so as to treat a non-tridiagonal-block Hamiltonian matrix. Let $A = \varepsilon I - H_C - \Sigma^{r,L} - \Sigma^{r,R}$, where $H_C$ is the Hamiltonian matrix of the channel region alone, and define $\delta_1^{(0)} = A_{0,0}$, $\delta_2^{(0)} = A_{1,1}$, $\alpha^{(0)} = A_{0,1}$, $\beta^{(0)} = A_{1,0}$. The algorithm consists in eliminating the layers from $1$ to $N_x$ with the formulas
\begin{align}
\delta_1^{(n)} &= \delta_1^{(n-1)} - \alpha^{(n-1)} \left[ \delta_2^{(n-1)} \right]^{-1} \beta^{(n-1)} \, , \nonumber \\
\delta_2^{(n)} &= A_{n+1,n+1} - \, A_{n+1,n} \left[ \delta_2^{(n-1)} \right]^{-1} A_{n,n+1} \, , \nonumber \\
\alpha^{(n)} &= -\alpha^{(n-1)} \left[ \delta_2^{(n-1)} \right]^{-1} A_{n,n+1} + A_{0,n+1} \, , \nonumber \\
\beta^{(n)} &= - A_{n+1,n} \left[ \delta_2^{(n-1)} \right]^{-1} \beta^{(n-1)} + A_{n+1,0}
\end{align}
for $n=1,\ldots,N_x$, where it is understood that $A_{0,N_x+1}=A_{N_x+1,0}^\dagger = 0$. At the end, the required matrix block of the Green function can be obtained as
\begin{equation}
G^r_{0,N_x+1} = - \left[ \delta_1^{(N_x)} \right]^{-1} \alpha^{(N_x)} \left\{ \delta_2^{(N_x)} - \beta^{(N_x)} \left[ \delta_1^{(N_x)} \right]^{-1} \alpha^{(N_x)} \right\}^{-1} .
\end{equation}

\end{document}